# Decay data of radionuclides along the valley of nuclear stability for astrophysical applications

CHECHEV Valery (В.П.契切夫) [1], HUANG Xiaolong (黄小龙) [2,*]

[1] V.G. Khlopin Radium Institute, 194021, Saint Petersburg, Russia

[2] China Institute of Atomic Energy, P.O. Box 275(41), Beijing 102413, China

**Abstract:** Several directions of the demand of decay data in nuclear astrophysics are discussed for radionuclides near the valley of nuclear stability. The current decay data evaluation results have been presented for a number of radionuclides of astrophysical interest. An extended list of such nuclides is offered for their nuclear characteristics to be further evaluated by the Decay Data Evaluation Project cooperation participants.

**Key Words**: Radionuclides, decay data, half-life, γ rays, nucleosynthesis, s-process.

## 1  Introduction

The nuclear astrophysics studies the nuclear processes in stars and cosmos and, first of all, nucleosynthesis. The latter includes Big Bang and stellar nucleosynthesis, supernova nucleosynthesis, nova nucleosynthesis and also cosmic ray spallation. To describe all these astrophysical processes at different temperatures and densities there are required from nuclear physics primarily the characteristics of nuclear reactions measured in laboratory experiments – cross sections, rates, branching, etc. However it turns out, radionuclide decay characteristics are important also in a number of requests of nuclear astrophysics. Here we consider several directions of the demand of decay characteristics in describing the nucleosynthesis processes for radionuclides near the valley of nuclear stability. Decay data of such radionuclides are being evaluated currently by the Decay Data Evaluation Project (DDEP) cooperation aimed to applied radionuclides [1-3]. In this paper an extended list of radionuclides of astrophysical interest has been proposed for evaluation of their decay data by the DDEP participants.

## 2  Half-lives of long-lived radionuclides

The first direction of the demand of decay characteristics is associated with half-lives ($T_{1/2}$) of long-lived radionuclides more than $10^8$ years ($^{40}$K, $^{87}$Rb, $^{176}$Lu, $^{187}$Re, $^{232}$Th, $^{235}$U, $^{238}$U) and also with half-lives of relatively short-lived radionuclides ($10^5$ yr < $T_{1/2}$ < $10^8$ yr), now extinct, decay of which provides information about the time interval during which the solidification of the planetary and meteoritic material occurs. Latter's are $^{26}$Al, $^{53}$Mn, $^{60}$Fe, $^{93}$Zr, $^{98}$Tc, $^{107}$Pd, $^{129}$I, $^{135}$Cs, $^{146}$Sm, $^{182}$Hf, $^{205}$Pb, $^{244}$Pu, and $^{247}$Cm [4].

---

[*] Corresponding author. e-mail: huang@ciae.ac.cn

A summary of recommended (evaluated) half-life values was given for the above 20 radionuclides in [4]. To date, due to the publication of more accurate measurements [5, 6], the recommended half-life values must be changed for $^{60}$Fe and $^{93}$Zr.

A new determination of the half-life of $^{60}$Fe using high precision measurements of the number of atoms and their activity in a sample containing over $10^{15}$ atoms of $^{60}$Fe [5] has given the $^{60}$Fe half-life value of $(2.62\pm0.04)\times10^6$ yr, significantly above the previously reported value of $(1.49\pm0.27)\times10^6$ yr. This new result of 2.62(4) Myr should be considered as the current recommended half-life of $^{60}$Fe for astrophysical applications.

For the half-life of $^{93}$Zr, we have adopted the DDEP evaluation by M.A. Kellett [7] which takes into account the recent measurement result of $(1.64\pm0.06)\times10^6$ yr [6]. Thus, the recommended $^{93}$Zr half-life is 1.61(6) Myr obtained as the weighted mean of [6, 8].

It should be noted that there is still only by one early half-life measurement for $^{98}$Tc, $^{107}$Pd and $^{205}$Pb. New measurements would be beneficial.

## 3 Radionuclides decay of which generates gamma rays observed by orbital detectors

The second direction of astrophysical interest includes decay data, mainly γ ray characteristics, for radionuclides decay of which generates gamma rays observed (or which can be observed) by orbital detectors. Radioactive isotopes are co-produced with stable isotopes in stellar interiors, supernovae, novae, and interstellar space. Stellar interiors are opaque, but expanding explosive sites of nucleosynthesis such as novae and supernovae are gamma ray transparent typically a few days to weeks after the explosion, thus allowing a direct gamma ray view at the nucleosynthesis site. Only a small number of isotopes produced in nucleosynthesis *in significant quantity* have half-lives sufficiently long to not have decayed already before transparency of the site: $^7$Be, $^{22}$Na, $^{26}$Al, $^{44}$Ti, $^{56}$Ni, $^{57}$Co, $^{60}$Fe [9]. Below we have given the recommended (evaluated) values of main decay characteristics for these radionuclides and their daughters (**Table 1**). γ ray characteristics are presented in the fourth column of the table: the evaluated energies (Eγ, keV) and absolute intensities per 100 decays (Iγ, %) of prominent γ rays.

**Table 1.** Recommended decay data for 10 radionuclides of astrophysical interest

| Nuclide | Half-Life | Decay | Eγ, keV / Iγ, % | Reference |
|---------|-----------|-------|-----------------|-----------|
| $^7$Be | 53.23(4) d [a] | $^7$Be→$^7$Li* | 477.6035(20) / 10.44(4) | [10] |
| $^{22}$Na | 2.6020(4) yr [a] | $^{22}$Na→$^{22}$Ne* | 1274.537(7) / 99.94(13) <br> (γ±) 511 / 180.7(2) | [11] |
| $^{26}$Al | 0.717(24) Myr | $^{26}$Al→$^{26}$Mg* | 1808.65(7) / 99.76(4) <br> (γ±) 511 / 163.5(4) | [12] |
| $^{44}$Ti/$^{44}$Sc [b] | 59.3(7) yr [c] | $^{44}$Ti→$^{44}$Sc*→$^{44}$Ca* | 78.3234(14) / 96.4(17) [d] <br> 67.8679(14) / 93.0(19) [d] <br> 1157.020(15) / 99.9(4) [d] <br> (γ±) 511 / 188(3) | [10] |
| $^{56}$Ni | 6.075(10) d | $^{56}$Ni→$^{56}$Co | 158.38(3) / 98.8(10) <br> 811.85(3) / 86.0(9) <br> 749.95(3) / 49.5(12) | [13] |
| $^{56}$Co | 77.236(26) d | $^{56}$Co→$^{56}$Fe* | 846.7638(19) / 99.9399(23) <br> 1238.2736(22) / 66.46(16) [e] <br> (γ±) 511 / 39.21(22) | [14] |
| $^{57}$Co | 271.80(4) d [a] | $^{57}$Co→$^{57}$Fe* | 122.06065(12) / 85.51(6) <br> 136.47356(29) / 10.71(15) <br> 14.41295(31) / 9.15(17) | [10] |
| $^{60}$Fe | 2.62(4) Myr | $^{60}$Fe→$^{60m}$Co→$^{60}$Co | 58.603(7) / 2.065(30) | [15] |
| $^{60}$Co | 5.2710(8) yr [a] | $^{60}$Co→$^{60}$Ni* | 1332.492(4) / 99.9826(6) <br> 1173.228(3) / 99.85(3) | [14] |

[a] The value revised by the authors of this work taking into account the 2012 information: [16] for $^7$Be and [17] for $^{22}$Na, $^{57}$Co and $^{60}$Co.
[b] $^{44}$Ti in equilibrium with $^{44}$Sc. The equilibrium is reached in 40 hours after synthesis of $^{44}$Ti.
[c] The value revised by the authors of this work taking into account the 2006 measurement [18].
[d] Re-evaluated by the authors of this work.
[e] The value revised by the authors of this work taking into account the 2008 measurement [19].

## 4 Radionuclides produced in nucleosynthesis s- and p- processes

The third field of decay data of astrophysical interest for radionuclides near the valley of nuclear stability relates to theoretical nuclear astrophysics, specifically to s- and p- processes of nucleosynthesis. Generally, there are three basic processes of stellar nucleosynthesis in which the chemical elements are produced: s-process (slow neutron capture alternating with beta decays occurring along the valley of nuclear stability), r-process (rapid multiple (repeated) neutron capture) and p-process (proton capture occurring in the limited field of mass numbers) [20]. Here we do not consider r-process as it occurs in the field of very short-lived highly neutron-rich nuclei.

Level schemes and decay data of so-called "key" nuclides produced in s-process are important for study of stellar interior where s-process occurs. Such nuclides have relatively long-lived (>1 ns) low-lying excited states including isomeric ones which are populated thermally in stellar conditions that leads to s-process branching. Accordingly, the track of s-process depends from the β-decay properties of these states and the temperature of stellar interior where s-process occurs. Therefore, based on the known relative abundances of the stable natural nuclides produced in the end of the s-process chain, we can estimate the temperature of s-process [21]. Decay data of one of the "key" nuclides produced in s-process – $^{79g,m}$Se – were evaluated recently in [4].

There are in total 30 branching points of the s-process: $^{63}$Ni, $^{64}$Cu, $^{79}$Se, $^{80}$Br, $^{81}$Kr, $^{85}$Kr, $^{93}$Zr, $^{99}$Tc, $^{107}$Pd, $^{113}$Cd, $^{129}$I, $^{134}$Cs, $^{135}$Cs, $^{147}$Pm, $^{151}$Sm, $^{152}$Eu, $^{153}$Gd, $^{154}$Eu, $^{155}$Eu, $^{160}$Tb, $^{163}$Dy, $^{163}$Ho, $^{170}$Tm, $^{171}$Tm, $^{176}$Lu, $^{182}$Ta, $^{192}$Ir, $^{193}$Pt, $^{204}$Tl, $^{205}$Pb. In the fourteen of them, the low-lying isomeric states occur: $^{79m}$Se, $^{80m}$Br, $^{81m}$Kr, $^{85m}$Kr, $^{99m}$Tc, $^{107m}$Pd, $^{113m}$Cd, $^{134m}$Cs, $^{135m}$Cs, $^{163m}$Ho, $^{176m}$Lu, $^{182m}$Ta, $^{192m}$Ir, $^{193m}$Pt.

The nucleosynthesis p-process also produces in significant quantity several interesting radionuclides with relatively long half-lives, including $^{92}$Nb ($T_{1/2} = 3.6 \times 10^7$ yr), $^{97}$Tc ($T_{1/2} = 2.6 \times 10^6$ yr), $^{98}$Tc ($T_{1/2} = 4.2 \times 10^6$ yr), and $^{146}$Sm ($T_{1/2} = 1.08 \times 10^8$ yr). Of them, the evidence for decay in meteorites was obtained so far only for $^{146}$Sm [22].

All the above radionuclides near the line of nuclear stability have astrophysical interest and can be offered for detailed evaluation of their decay characteristics by the DDEP cooperation participants.

## 5 Conclusion

High-quality detailed evaluated decay data are required in many applications uncluding nuclear astrophysics. We have presented the current evaluated decay data for several radionuclides of astrophysical interest and offered the extended list of such radionuclides for future evaluation.

As members of the DDEP cooperation, we would like to encourage other scientists to collaborate with us and help us expand the Decay Data Evaluation Project to evaluate and, if necessary, to measure many decay data of astrophysical interest.